\documentclass[manuscript, nonacm]{acmart}
\AtBeginDocument{%
  }

\setcopyright{none}
\renewcommand\footnotetextcopyrightpermission[1]{}

\usepackage{listings}
\usepackage{tabularx}
\usepackage{enumitem}
\usepackage[most]{tcolorbox} 
\tcbset{
  colback=black!3,
  colframe=black!40,
  boxrule=0.4pt,
  arc=1pt,
  left=8pt, right=8pt, top=8pt, bottom=8pt
}

\begin{document}

\title{Designing Mental-Health Chatbots for Indian Adolescents: Mixed-Methods Evidence, a Boundary-Object Lens, and a Design-Tensions Framework}

\author{Neil K. R. Sehgal}
\email{nsehgal@seas.upenn.edu}
\affiliation{%
  \institution{University of Pennsylvania}
  \city{Philadelphia}
  \state{Pennsylvania}
  \country{USA}
}

\author{Hita Kambhamettu}
\email{hitakam@seas.upenn.edu}
\affiliation{%
  \institution{University of Pennsylvania}
  \city{Philadelphia}
  \state{Pennsylvania}
  \country{USA}
}

\author{Sai Preethi Matam}
\email{saipreethimatam@gmail.com}
\affiliation{%
  \institution{Mamata Academy  of Medical Sciences}
  \city{Hyderabad}
  \state{Telengana}
  \country{India}
}

\author{Lyle Ungar}
\email{ungar@cis.upenn.edu}
\affiliation{%
  \institution{University of Pennsylvania}
  \city{Philadelphia}
  \state{Pennsylvania}
  \country{USA}
}

\author{Sharath Chandra Guntuku}
\email{sharathg@seas.upenn.edu}
\affiliation{%
  \institution{University of Pennsylvania}
  \city{Philadelphia}
  \state{Pennsylvania}
  \country{USA}
}

\renewcommand{\shortauthors}{Sehgal et al.}

\begin{abstract}

Mental health challenges among Indian adolescents are shaped by unique cultural and systemic barriers, including high social stigma and limited professional support. We report a mixed-methods study of Indian adolescents (survey n=362; interviews n=14) examining how they navigate mental-health challenges and engage with digital tools. Quantitative results highlight low self-stigma but significant social stigma, a preference for text over voice interactions, and low utilization of mental health apps but high smartphone access. Our qualitative findings reveal that while adolescents value privacy, emotional support, and localized content in mental health tools, existing chatbots lack personalization and cultural relevance. We contribute (1) a Design-Tensions framework; (2) an artifact-level probe; and (3) a boundary-objects account that specifies how chatbots mediate adolescents, peers, families, and services. This work advances culturally sensitive chatbot design by centering on underrepresented populations, addressing critical gaps in accessibility and support for adolescents in India.

\end{abstract}




\maketitle

\section{Introduction}
Adolescent mental health (MH) is a global concern. In low- and middle-income countries (LMICs) like India—home to 20\% of the world’s under-25s—systemic and societal barriers intensify stress, anxiety, and depression \cite{patel2007mental, Naveed2020PrevalenceOC, gallup2025}. India faces the highest treatment gap across Asia, with 95\% of adults lacking access to mental healthcare \cite{murthy2017national, Naveed2020PrevalenceOC}.

AI-driven MH tools offer scalable, anonymous support, yet most are trained on Western data and design assumptions, potentially limiting cultural fit in LMICs \cite{kozelka2021advancing, cho-etal-2023-integrative, rai2024key}. For Indian adolescents navigating stigma, resource gaps, and socio-emotional needs, Western-centric systems may miss local realities. We frame chatbots not only as informational tools but as sociotechnical boundary objects—artifacts that mediate relationships among adolescents, peers, families, and systems of care \cite{lutters2002achieving}. This lens foregrounds trust, stigma management, and cultural fit, positioning chatbots as complements—not replacements—to human care.


To advance this agenda, we report a mixed-methods study of Indian adolescents (survey n=362, interviews n=14) and address three research questions:

\begin{description}
    \item[RQ1:] \textbf{How do Indian adolescents perceive and navigate mental health challenges within societal and cultural contexts?}
    \item[RQ2:] \textbf{What are the experiences and challenges Indian adolescents face when using current digital tools to seek mental health support?}
    \item[RQ3:] \textbf{How can chatbots be designed to better support the mental health needs of Indian adolescents?}
\end{description}

We additionally ask a pragmatic RQ4: \textit{How should adolescent MH chatbots in India align with national infrastructure (Tele‑MANAS) \cite{ahmed2022tele} and data protection (Digital Personal Data Protection, DPDP 2023) \cite{jha2024dpdp} to be deployable at scale?}

We find persistent tensions: adolescents are open to digital tools and show low self-stigma, yet face pervasive social stigma and rarely use existing apps. Chatbots can curate concise, emotionally attuned guidance, but current tools often lack cultural fit, personalization, and privacy.

From this, we make four contributions to HCI and CHI’s global health and HCI4D communities:

\begin{description}
    \item[1] \textbf{Empirical Insights}  A mixed-methods account of Indian adolescents’ (n= 362 / 14) mental health needs and technology practices, surfacing barriers (e.g., stigma, low app use) and preferences (e.g., text over voice, privacy, localized support).
    \item[2] \textbf{Design-tensions framework} A set of interaction principles and safeguards, tied to WHO LLM guidance ~\cite{world2024releases} and NICE ESF and NHS DTAC ~\cite{national2019evidence,daniel2022toolkit}, for adolescent mental health chatbots in LMICs, derived from survey patterns and interview feedback on privacy, memory, guidance styles, modality, curation, and persona fit.
    \item[3] \textbf{Theoretical framing} A conceptualization of chatbots as sociotechnical boundary objects that complement, rather than replace, human care.
    \item[4] \textbf{Artifact-level probe} A lightweight prototype and prompting pattern evaluated via role-play, with implications for mixed-initiative, supportive interactions.
\end{description}

We also discuss infrastructure alignment, showing hand-offs to Tele-MANAS and privacy controls under DPDP. 

\section{Related Work}

\subsection{Chatbots for Adolescent Mental Health in LMICs}

The scalability, anonymity, and always-on nature of chatbots have made them an increasingly popular tool in global MH initiatives targeting youth. Particularly in LMICs, where access to formal mental healthcare remains limited, chatbots offer a promising pathway to expand reach and reduce stigma. Large-scale implementations include UNICEF’s Pode Falar (Brazil), WHO’s STARS (South Africa, Nepal, Pakistan), E.V.A. (Peru, for HIV-positive adolescents), and SuSastho.AI (Bangladesh) \cite{da2024escuta, who, rupani2025like, al2024ai}.
These efforts highlight chatbots’ utility, yet design often remains under-theorized for cultural specificity, language diversity, and youth-centered interaction—especially in India.


\subsection{Cultural Contexts and Mental Health Technologies}

Help-seeking, emotion regulation, and support norms are culturally patterned \cite{rai2024key, krendl2020countries}. Culturally adapted interventions can be up to four times more effective \cite{griner2006culturally}, yet many AI tools reflect WEIRD populations \cite{kozelka2021advancing, cho-etal-2023-integrative}. In India, stigma and norms around disclosure shape adolescent preferences, and mismatches may reduce engagement and relevance \cite{murthy2017national}.


\subsection{Designing for Adolescents in South Asia}

HCI work in South Asia highlights how sociopolitical context, gender, and class shape mental-health needs and technology fit: e.g., digital MH among working-class Indian women \cite{reen2022improving}, support structures for Kashmiri youth amid conflict \cite{wani2024unrest}, and chatbot interventions for Bangladeshi adolescents \cite{rahman2021adolescentbot}. These projects argue for co-design, contextual sensitivity, and infrastructural awareness. Despite India’s large youth population, they remain underrepresented in chatbot-specific research. Our study centers their needs to inform culturally grounded chatbot design.

\subsection{Chatbots as Boundary Objects}

We treat adolescent MH chatbots as \textit{boundary objects}—artifacts accommodating multiple meanings, aka interpretive flexibility, across communities (friend, guide, tool) while bridging social worlds (e.g. adolescents, families, schools, and services)  \cite{lutters2002achieving,star1989institutional}. This interpretive flexibility is especially relevant in stigmatized domains like adolescent MH, where direct engagement with formal systems may be avoided \cite{yoo2024missed}. Prior work has examined boundary objects in digital health infrastructures \cite{zhou2011cpoe}, but less attention has been paid to adolescent-facing chatbots in complex LMIC settings and how they mediate stigma, trust, and handoffs to human care. 

Across these strands, we address two needs: (1) translating cultural and developmental considerations into interaction-level design tensions and principles for adolescent chatbots in India, and (2) articulating infrastructure-aware integration patterns (school, family, helplines) that treat chatbots as complements, not replacements, to human care.

\section{Methods}
\textbf{Study Design}
We employed a mixed-methods design with two phases: (1) a quantitative survey to capture broad trends and (2) qualitative interviews to gather contextual insights. This triangulation supported comprehensive recommendations for culturally inclusive digital MH tools. Please see the Appendix for an extended methodology.

\textbf{Participant Recruitment}
Participants were recruited via social media and outreach to school administrators, targeting diverse geographic and socioeconomic backgrounds. Survey respondents could volunteer for follow-up interviews. All participants gave informed consent; the study received institutional ethics approval. Small gift cards were provided as appreciation.

\textbf{Phase 1: Survey}
The anonymous survey examined MH experiences, barriers to seeking support, and preferences for digital tools. It included Likert, multiple-choice, and open-text items, partially adapted from validated surveys/scales and refined with Indian counselors \cite{vogel2006measuring, murthy2017national, mental_health_survey_2022, lowe2005detecting}. Responses were collected in English via Qualtrics.

\textbf{Phase 2: Qualitative Interviews}

Semi-structured interviews with 14 participants explored survey themes through role-play and discussion. In the role-play, participants chose one of six counselor-informed scenarios (Appendix). They completed two tasks:

\begin{itemize}
    \item \textbf{Searching for Information Online} Participants sought solutions to the scenario via any platform (e.g., Google, Instagram, Reddit) while screen-sharing; chatbots were excluded.

    \item \textbf{Chatbot Interaction}: They then used a prototype chatbot (built on GPT-4o-mini) designed to provide empathy and practical suggestions. The chatbot was constrained by a predefined prompt (Appendix) and structured to give concise, supportive responses. 
    The bot never offered diagnosis/treatment, in line with WHO LMM guidance. Session logs were set to zero‑retention during the study.
\end{itemize}

\begin{figure}[]
\centering
\includegraphics[width=.6\textwidth]{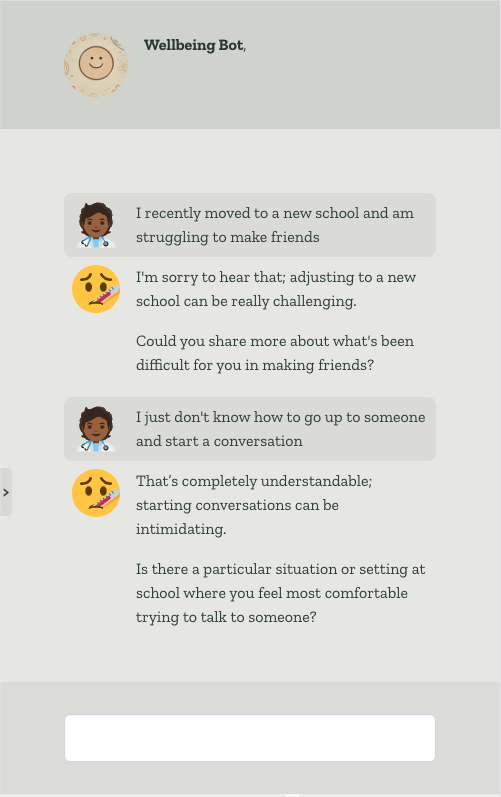}
\caption{The chatbot interface. Interviewees could access the chatbot on their phone, tablet, or laptop and interacted via text. The chatbot was programmed to always give short responses and ask one question per turn.}
\end{figure}

Interviews were conducted in English via Zoom, lasting 45–60 minutes.

\textbf{Data Analysis}

We used descriptive statistics and Pearson ${\chi}^2$ tests (p<.01) for age comparisons (<18 vs. $geq$18). Thematic analysis involved open coding by one author, reviewed and refined collaboratively with a second. Codes not relevant to digital tool design were discarded. To ensure validity, the team engaged in iterative, collaborative discussions at each stage of analysis, refining codes to maintain accuracy and consistency. This approach emphasized contextual depth and shared interpretation rather than numeric agreement metrics \cite{mcdonald2019reliability, kambhamettu2024explainable, sehgal2025paldesigningconversationalagents}.



\section{Phase 1: Survey Findings}

The survey included responses from 362 adolescents across India. Notable findings include:

\textbf{Limited Uptake:} A significant majority (71.3\%) had never accessed MH services. Of those who had, 44.2\% consulted school counselors, while only 5.7\% used MH apps, highlighting a large treatment gap.

\textbf{Low Self-Stigma but High Social Stigma: }Only 11.9\% of participants reported feelings of inadequacy when considering therapy (self-stigma). However, 39.2\% worried about negative societal perceptions if they sought professional help (social stigma), underscoring the cultural barriers to formal care.

\textbf{Mental Health Chatbots: Low Usage but High Perceived Helpfulness:} While only 27.4\% had used chatbots to talk about their feelings or problems, 79.3\% of these users reported finding them helpful. This highlights their potential, despite low adoption rates, to address gaps in traditional care.

\textbf{Skewed Preferences for Text-Based Interactions:} A majority (68.2\%) preferred text-based communication with chatbots, compared to 6\% favoring voice.

A summary of key survey results is provided in Table~\ref{tab:survey_data}.

\begin{table}[t]
\caption{Survey Data Summary}
\label{tab:survey_data}
\centering
\small
\begin{tabular}{p{0.35\linewidth} p{0.6\linewidth}}
\toprule
\textbf{Characteristic} & \textbf{Details} \\
\midrule
\multicolumn{2}{l}{\textbf{Demographic Characteristics}} \\
Participants  & 362 students (27\% Class 9, 44\% first-year university, 26\% second-year university) \\
Age           & 12--23 years (mean: 17.9 years (std: 2.5), under 18: 32\% ) \\
Gender        & 67\% female, 30\% male, 3\% undisclosed \\
Religion      & 85\% Hindu \\
\midrule
\multicolumn{2}{l}{\textbf{Mental Health Challenges}} \\
Loneliness    & Over the past 12 months, 37\% felt lonely "sometimes", 25\% "most of the time" , 5\% "always" \\
Prolonged sadness &  Over the past 12 months, 35\% experienced sadness/hopelessness for >2 weeks \\
Sleep disruption  & Over the past 12 months, 33\% reported stress-related sleep disruption "occasionally", 16\% "most of the time", 4\% "always" \\
Pressure for good grades & 34\% felt "a lot", 45\% felt "some" pressure \\
Social pressure to fit in  & 25\% experienced "a lot", 44\% "some" pressure \\
\midrule
\multicolumn{2}{l}{\textbf{Past Experience with Mental Health Support}} \\
Never accessed services & 71.3\% \\
Common resources used by those who had accessed services & school counselors (44\%), peer support groups (29\%), private therapists (21\%), online counseling (8\%), mental health apps (6\%) \\
\midrule
\multicolumn{2}{l}{\textbf{Attitudes Towards Mental Health}} \\
Supportive communities & 54\% community, 64\% parents, 70\% friends deemed supportive/very supportive \\
Mental health discussions & Rarely/never with family (63\%); frequently/occasionally with friends (59\%) \\
\midrule
\multicolumn{2}{l}{\textbf{Barriers to Mental Health Support}} \\
Self-stigma   & If they sought therapy, 12\% would feel inadequate, 14\% self-dissatisfied, 12\% inferior \\
Social stigma & If they sought therapy, 41\% concerned about negative reactions from others \\
\midrule
\multicolumn{2}{l}{\textbf{Digital Resource Use and Preferences}} \\
Technology access & 82\% owned personal smartphones, 94\% frequently used the internet \\
Chatbot usage     & 83\% had used chatbots; 27\% for personal issues (79\% found these personal conversations helpful) \\
Desired features  & Anonymity (54\%), quick responses (61\%), access to resources (22\%), peer support (23\%) \\
\midrule
\multicolumn{2}{l}{\textbf{Chatbot Concerns and Desired Improvements}} \\
Key concerns      & Privacy (68\%), lack of personal connection (48\%), cost (37\%), reliability (35\%) \\
Desired features  & Coping strategies (72\%), personalized recommendations (69\%), regular check-ins (51\%) \\
Preferred communication & Text (68\%), voice (6\%), both (26\%) \\
Language preference     & 94\% English \\
\bottomrule
\end{tabular}
\end{table}

\pagebreak

Many survey responses were similar across age groups (Under 18 vs 18 and up). For instance, 69.0\% of those under 18, and 75.6\% of those 18+ had never accessed mental-health services(p=.22). However, several differences did emerge (Table~\ref{tab:age_single}, \ref{tab:age_multi}).

\textbf{Parental nexus intensifies below 18.} Under-18 respondents both receive more daily affection and rely more on parents for academic advice.

\textbf{Peer‑based mental‑health talk matures later.} 18+ respondents discuss MH with friends far more often, suggesting chatbots for younger teens may need built‑in primers on how to start such conversations.

\textbf{Help‑seeking channels diverge.} Under‑18 respondents turn to school counselors; 18+ respondents cite cost as a key barrier. This indicates financial‐access design features (price transparency, subsidized referrals) are especially critical for university‑age users.

\textbf{Everyday negative affect is still higher among younger respondents.} Frequent feelings of being “put down” remain a salient stressor, underscoring the importance of anti‑bullying and self‑esteem modules for younger cohorts.

\begin{table}[ht]
\caption{Largest age–group differences on single‑choice items ($p<.01$)}
\label{tab:age_single}
\centering
\small
\begin{tabular}{p{0.35\linewidth} p{0.23\linewidth} rr c}
\toprule
\textbf{Survey item (abridged)} & \textbf{Response category\,$^{\dagger}$} &
\textbf{Under 18 (\%)} & \textbf{18 + (\%)} & Cramer's $V$ \\
\midrule
People \emph{put you down} & Almost/Every day & 34 & 26 & .25 \\
Get a \emph{hug or kiss from parents} & Almost/Every day & 53 & 30 & .24 \\
Discuss mental health with \emph{friends} & Frequently/Occasionally & 47 & 64 & .22 \\
Ask parents for \emph{homework help} & Almost/Every day & 40 & 20 & .22 \\
Wish you had more \emph{good friends} & Almost/Every day & 26 & 36 & .20 \\
Get excited by \emph{something you studied in school} & Almost/Every day & 27 & 45 & .19 \\
\bottomrule
\multicolumn{5}{p{0.93\linewidth}}{\footnotesize
$^{\dagger}$The response category that most clearly differentiates the two age groups (others collapsed). 
Percentages are row‑normalised within age group (Under 18 $n=116$, 18+ $n=246$).}
\end{tabular}
\end{table}

\begin{table}[ht]
\caption{Largest age–group differences on multi‑select items ($p<.01$)}
\label{tab:age_multi}
\centering
\small
\begin{tabular}{p{0.25\linewidth} p{0.38\linewidth} rr c}
\toprule
\textbf{Survey item} & \textbf{Response option} &
\textbf{Under 18 (\%)} & \textbf{18 + (\%)} & $\phi$ \\
\midrule
Accessed the following mental health service & School counselor & 26 & 7 & .262 \\
Concerns around digital mental health services & Cost barrier & 27 & 42 & .137 \\
\bottomrule
\multicolumn{5}{p{0.93\linewidth}}{\footnotesize
Percentages indicate the share of each age group selecting the option in a “check‑all‑that‑apply” format. 
Effect size is the $\phi$ coefficient from a 2 × 2 $\chi$\textsuperscript{2} test (Under 18 $n=116$, 18+ $n=246$).}
\end{tabular}
\end{table}

\section{Phase 2: Interview Findings}

We interviewed 14 participants ranging in age from 14 to 20. The majority (79\%) of participants identified as female, reflecting the makeup of survey participants. All but one participant was age 18 years or over. A plurality (n=5) chose the roleplay scenario of struggling to get into university. Appendix Table 1 displays participant demographics.

\subsection{How Indian Adolescents Perceive and Navigate Mental Health}

\subsubsection{Stigma and fear of judgement} \label{stigma}
As in the survey, many participants (N=11) cited social stigma as a major barrier to seeking MH support. While most survey respondents noted supportive family attitudes, they rarely discussed MH with them. Only one interviewee (P13) felt comfortable sharing with parents, while others feared judgment or disapproval. For example, P9 avoided official support to keep it off record, knowing their parents wouldn’t approve. This highlights the need for anonymous digital tools that limit data collection and give users control. Participants were also reluctant to confide in friends, and generational differences added challenges—older relatives often dismissed MH as weakness or money-making. Rural participants described further difficulties, with P5 noting, “Mental health problems are not considered problems. It’s just overthinking.”

\subsubsection{Additional Barriers} \label{barriers}
In line with our survey findings on low uptake of counseling, only one participant had sought counseling in the past, and few knew others who had. Beyond stigma, barriers included uncertainty about where to turn, lack of teacher support, cost, and dissatisfaction with services. P5 recalled struggling in 10th standard without anyone to reach out to, and said teachers dismissed her stress as exam-related. P1 noted therapy could be too expensive for students. P10, who had tried counseling, felt misunderstood. P14, prescribed anxiety medication, had consulted only a general physician, saying, “I’m yet to consult a psychiatrist. I wanted to deal with that [the physical manifestation] first. I know there's an underlying cause.”

\subsubsection{Reliance on friends for support} \label{friends}
Our survey found high levels of supportive attitudes from friends. Similarly, interviewees cited friends as their most common source (N=8) of emotional support, offering a judgment-free space for discussing MH concerns. P1 shared, “Talking to friends is how we usually deal with MH problems. I don’t know anyone who has consulted a therapist.” However, participants acknowledged not all of their friends felt comfortable sharing their problems, and others mentioned limitations of relying solely on friends. P1 recounted feeling hesitant towards going to their friends again after a recent issue they shared, stating “I felt something like being judged.” This suggests a gap where digital tools could complement informal peer networks by providing a judgment-free, always-available source of support.

\subsubsection{Role of Self-Help and Informational Gaps}
When traditional support systems were unavailable, participants turned to self-directed strategies, including hobbies, sleeping, journaling, and online platforms. P3 noted, “If I have no one near me, I just take my phone, open Google, and type my problem to seek answers.” However, many expressed dissatisfaction with existing digital resources, citing issues of reliability, lack of personalization, and overwhelming amounts of information. 

\subsection{Experiences with Current Technologies Around Mental Health}

\subsubsection{Navigating Search Engines}
Participants often turned to Google for MH advice because it was familiar and accessible, but struggled with overwhelming, unreliable results. P11 noted, “I had to search a bit after typing the topic. It wasn’t easy to find what I wanted.” P12 stated, "I just want a gist of what I need to know, not long articles." Many criticized search engines for excessive, uncurated information or paywalls. P4 remarked, "It’s all over the place—there are no curated spaces for specific mental health information" and noted that searching while anxious or depressed would make it even harder: “I wouldn’t be very proactive in my search…I would already be anxious, and it would be a lot of work.” P13 expressed disliked Google's AI overviews: "I asked 3 different questions, but the [Google AI] answers remained pretty much the same...very repetitive...very broad and they lack emotional support."

\subsubsection{Lack of Relevance} \label{relevance}
The lack of culturally relevant, trustworthy MH information was a recurring frustration for participants. P6 shared, “It was a bit helpful but not completely helpful for us.” P9 stated, “I found a helpline, but it was for California—how is that helpful to me?” P4 explained, “As an Indian, I'm experiencing something extremely different from something an American or an Australian is experiencing when it comes to not just exams, but also things to do with relationships and financial constraints”. This highlights the need for digital tools that curate locally relevant information and connect users to appropriate resources.

\subsubsection{Role of Social Media and Community Platforms}
While the majority of survey respondents mentioned an interest in self-help tips, fewer interviewees shared this. Instead, many interviewees discussed wanting to read about others' personal experiences. P3 emphasized, "I want reassurance that someone went through this and succeeded." Still some disagreed. P4 commented, "There’s very little that personal testimonies alone can do to help." Moreover, while platforms like Quora and Reddit provided opportunities to seek advice and share experiences, they were often viewed as not specific enough. P3 shared, “Quora feels less trustworthy because everyone’s experience is different.”

\subsubsection{Existing Chatbots}
All participants had used chatbots for school, but only four for personal or MH conversations, mirroring the survey distribution. Those who had personal conversations with chatbots valued accessibility and anonymity, seeing them as viable alternatives when feeling judged by others. P3 explained, “When I feel judged by my friends or family, I’d rather use a chatbot.” P5 shared, “I used ChatGPT to calm myself when I didn’t want to talk to anyone. It gave advice and reminded me to take deep breaths.” Participants liked chatbots' direct advice. P3 remarked, "ChatGPT is better than Google because it tells you how to act, not just reviews or random links." Still, some found responses generic. P10 said, “ChatGPT gives general bullet points, but doesn’t feel tailored to me.” Others avoided personal conversations, citing lack of emotional understanding (P4: “It’s a machine. It doesn’t understand feelings”), formality (P2: “Whatsapp or ChatGPT feel more like an official thing, not a personal chatbot”), or privacy concerns (P12 quit registration after being asked for a phone number).

\subsection{Desired Improvements for Digital Mental Health Tools}
In addition to understanding participant’s attitudes towards current technologies, we sought their recommendations for improvement. Based on interaction with our custom chatbot, participants proposed eight key improvements.

\subsubsection{Integration of External Resources}
To overcome the lack of relevance described in \ref{relevance}, participants recommended integrating localized resources. Two participants wanted chatbots to provide direct therapist contact when necessary. Five participants suggested chatbots link to external articles, social media posts, or videos showing other people have similar experiences to them. P7 explained, “like an article showing the problems in real life, and the person who succeeded with them.” Other participants suggested chatbots could provide videos, articles, or statistics from authoritative sources to back up its suggestions. However, participants maintained chatbots should mainly use such resources as citations. P12 explained, “I don't want the chatbot to provide me another link to go look at...they can probably make a gist out of it.” Another participant suggested "when it gives those suggestions like join a club, if there was like online clubs that it could link to, I think that would be really great" (P9).

\subsubsection{Improved Anonymity and Privacy Features}
To address the stigma described in \ref{stigma}, participants emphasized the importance of robust privacy features. P12 explained, “The idea of a chatbot is you know you are able to talk to somebody anonymously, and nothing is being recorded...if it takes your phone number and email address, that feeling of being anonymous goes away.” Similarly, P9 stated “I don't know how far my trust goes, like I wouldn't want it to ask what my name is.” Suggested features included password protection and the ability to delete past conversations. P3 explained, “If there is some password option...I'd feel pretty confidential about it.” Still, some felt a chatbot could never be truly private. P10 stated “I think there is no version of a chatbot which would feel my data is confidential because maybe somewhere it is being used.” These preferences are consistent with DPDP 2023, notably verifiable parental consent for under 18s, and argue for consent‑minimizing designs (e.g. guest mode, local transcript export+delete).

\subsubsection{Memory \& History}
All 14 participants wanted the chatbot to store previous conversations in some manner. P2 felt this would make interactions more personal, while others raised concerns. P4 stated “I wouldn't want it to know a lot of the things that I'm going through but also its more convenient to just build on things I’ve said. I wouldn't mind it having all of my information as long as there is a clear kind of privacy information.” P9 suggested asking permission before saving. Participants also wanted the ability to read past conversations. P6 stated “It would be helpful to go back and look after them again and again, and keep on implementing a few of the steps given by the chatbot.” However, as noted in \ref{stigma}, privacy concerns must be addressed to encourage trust. P5 mentioned: “sometimes I delete the conversations from ChatGPT manually. I feel embarrassed that I shared these kinds of things.” Others proposed exporting transcripts locally before deleting them from the platform. We therefore recommend a zero‑retention default with explicit, reversible `Save', local export, and a passcode—consistent with DPDP 2023 consent and withdrawal principles.

\subsubsection{Multi-Modal Interaction Options}
Only one participant expressed a clear desire for voice communication over text. However, some participants expressed interest in having both text and voice interaction options for different contexts. P8 remarked, "Sometimes I just want to speak instead of typing—it would make it more accessible." Similarly, P2 stated “I think texting anywhere is fine even in public…but voice chat only in my room.” P13 felt it could sometimes be difficult to express MH concerns in words: "most of the people who are actually dealing with mental issues...one of their core problems is they are not able to express in words with their mouth. So texting at the starting point..would be better." Participants also wanted tools to cater to diverse linguistic backgrounds for better accessibility and comfort.

\subsubsection{Balanced Response Length}
Participants preferred responses that were concise but provided sufficient detail to be helpful. Four participants felt our chatbot’s responses were too short. “I wish it gave me a little more, but not longer to the point where there's multiple paragraphs.” (P4). However, others disagreed and preferred the length, especially compared to ChatGPT. P13 explained, "the lengthier its responses are, the more anxious it makes a person to read through all of it." Seven participants expressed being particularly impressed by how "interactive" the chatbot felt. 

\subsubsection{Enhanced Personalization}
Participants generally liked the chatbot’s many questions, which felt more personalized than ChatGPT or Google. However, three wanted even more. P3 shared, “I would include more questions to assess the situation better on the first interaction.” Participants recommended incorporating mixed-initiative features, such as structured options to help guide the interaction when they were unsure of their needs. P4 suggested the bot offer "trajectories" to guide the conversation: "I wish the bot would respond like, 'ok, I understand this is your concern...Do you want me to talk about the academic patterns that you can change? Or do you want me to talk about your personal feelings?'" P4 noted benefits of past experience with AI chatbots, and knowing how to actively drive the conversation. “I was giving it context. But that's because I'm familiar with ChatGPT and interactive AI works in that sense…I know a lot of people don't have that information. So for it to be asking questions to you first, to gain a understanding of who you are would be a huge improvement (P4). 

\subsubsection{Emotional Support Features}
To supplement the peer networks described in \ref{friends}, participants desired chatbots to provide reassurance and nonjudgmental emotional support during stressful moments. P3 shared, “ChatGPT would not assure me it was okay to be this way. But when I asked a question, this chatbot assured me it was okay, it made me feel it was okay.” Participants appreciated the bot gave direct advice and encouraged self reflection. P5 shared, “this chatbot was different from ChatGPT. It was trying to push me to think about myself, think about my scenarios, rather than giving just direct answers. I need suggestions…take a deep breath, drink water. But before that I'll prefer [this] kind of engagement.” However, one participant expressed discomfort with self-reflection. P12 shared, “I want to go to a chatbot or anybody who I'm talking to, for comforting words. I want to be heard, I want to be comforted. But when I get asked questions about certain things that I even haven't thought of, it's forcing me to take more stress...It might not help.”

\subsubsection{Customizable Personality and Tone}
Some participants wanted to select a chatbot personality based on their comfort level. P2 suggested, "People should get to choose the personality they want. Some might feel comfortable with a friend, but some might feel comfortable with an expert or a mentor." 
P2 felt, “how they talk to you, how they present it, not the information itself, but the way the conversation goes is different [across personalities].”
In addition, some participants felt the bot’s current personality was that of a friend (n=4), advisor (n=3), mental health expert (n=1), or peer (n=1). 
Two participants felt the bot had no distinguishable personality. One participant emphasized keeping the conversation informal for comfort, while another participant shared that they felt a chatbot “could never judge” (P7). 
One participant expressed distaste for anthropomorphization, suggesting the chatbot should not have a name or avatar.

\section{Discussion}

This study reveals how Indian adolescents navigate mental health challenges, and how digital tools like chatbots may support them. In this section, we translate our empirical findings into design knowledge, framing implications for HCI scholarship and practice.

\subsection{Design Tensions for Indian Adolescent Mental Health Chatbots}
Building on our mixed-methods findings, we distill a set of design tensions that capture the key tradeoffs adolescents may face when engaging with digital MH tools in India. Rather than universal best practices, these tensions reflect situated struggles between privacy and continuity, guidance and autonomy, cultural fit and generic content that emerged consistently across survey and interview data. By articulating each tension alongside evidence, concrete implications, and proposed metrics, we offer a framework that translates adolescents’ lived experiences into actionable design principles. These metrics translate the framework into a shareable rubric that future designers, NGOs, or evaluators could adopt when assessing adolescent mental-health chatbots.

\begin{table*}[t]
\small
\centering
\caption{Design tensions for Indian adolescent mental-health chatbots, distilled from survey and interview data. Each tension pairs user evidence with concrete implications, example implementation patterns, and possible metrics. These metrics could be operationalized into a shareable rubric for evaluating or comparing systems.}
\label{tab:design-tensions}
\begin{tabularx}{\textwidth}{
>{\raggedright\arraybackslash}X
>{\raggedright\arraybackslash}X
>{\raggedright\arraybackslash}X
>{\raggedright\arraybackslash}X
>{\raggedright\arraybackslash}X}
\toprule
\textbf{Tension} & \textbf{What participants told us} & \textbf{Design implication} & \textbf{Example implementation} & \textbf{Possible metrics} \\
\midrule
\textbf{Privacy \(\leftrightarrow\) Memory} &
Anonymity is essential; phone/email collection deters use. Yet users want continuity, the ability to revisit chats, and selective history. &
Default to privacy-first storage with user control over retention. Make saving explicit and reversible; allow local export + chat deletion; optional passcode. &
“Save this conversation?” toggle at end of session; \emph{Zero-retention} mode; \emph{Chat Lock} gate; \emph{Export as PDF} before delete. &
\% of users enabling “Save”; \% deleting/exporting transcripts; user-reported trust in privacy (Likert). \\
\addlinespace
\textbf{Direct Advice \(\leftrightarrow\) Reflective Prompting} &
Some want concrete next steps and reassurance; others find probing questions stressful. &
Offer user-selectable response modes with mid-chat switching. &
Mode chips: \emph{Comfort} / \emph{Practical Steps} / \emph{Reflection}; “Switch mode” quick action in header. &
\% of sessions with mode switching; average duration of reflective sequences; change in perceived stress/anxiety (self-report). \\
\addlinespace
\textbf{Mixed-Initiative Guidance \(\leftrightarrow\) User Control} &
Novices want guided trajectories; experienced users prefer freeform control. &
Provide light-weight, skippable flows plus free-text at all times. &
Starter cards (e.g., \emph{Study Stress}, \emph{Friend Conflict}, \emph{Sleep}); “I’ll type my own” escape on every step. &
\% choosing guided vs.\ freeform start; dropout rate during guided flows; mean session length. \\
\addlinespace
\textbf{Local Curation \(\leftrightarrow\) Link-Dumping} &
Generic/foreign resources feel irrelevant; users prefer in-chat gists and India-specific options. &
Summarize sources in-chat; attach trusted, geo-scoped resources only when helpful. &
Inline 2–3 sentence gist + \emph{Show sources}; India helplines/therapist directories with region filters. &
Click-through rate on local resources; \% of resources rated “relevant”; crisis handoff success rate. \\
\addlinespace
\textbf{Text-First \(\leftrightarrow\) Voice-Optional} &
Strong preference for text; voice is situational in shared homes. &
Default to text; provide optional push-to-talk with clear privacy cues. &
Mic button with “no audio stored”; automatic default to text. &
\% sessions using voice; reported comfort with modality; error rate in speech input. \\
\addlinespace
\textbf{Persona Fit \(\leftrightarrow\) Tone Discipline} &
Some prefer friend-like tone; others want mentor/expert; some dislike anthropomorphism. &
Configurable personas with clear scope boundaries and non-deceptive presentation. &
Persona picker (\emph{Friend} / \emph{Mentor} / \emph{Expert}); optional minimal avatarless style. &
Distribution of persona selections; persona fit satisfaction score (Likert); frequency of persona switching. \\
\addlinespace
\textbf{Concise \(\leftrightarrow\) Sufficient Detail} &
Long replies raise anxiety; ultra-short replies feel thin. &
Default to concise responses with progressive disclosure. &
3–4 sentence answer + “Tell me more” expander; bullet next-steps + optional deeper dive. &
Median reply length; \% of users clicking “Tell me more”; satisfaction with clarity (Likert). \\
\bottomrule
\end{tabularx}
\end{table*}

\textbf{Reframing Chatbots as Sociotechnical Boundary Objects}
We frame MH chatbots as boundary objects: artifacts that mediate between adolescents, caregivers, peers, and institutional systems while accommodating diverse interpretations and uses. For adolescents navigating stigma, chatbots can function as confidants, self-reflection tools, and bridges to broader support networks. Their interpretive flexibility allows them to be perceived as a friend, guide, or expert depending on the user's emotional needs and context \cite{yoo2024missed}. 

Participants' feedback underscores the tensions in this boundary role: many adolescents desired emotional resonance and reassurance—traits typically associated with human care—yet some remained skeptical of AI’s capacity for empathy. To resolve this tension, designers must clarify chatbots’ roles within sociotechnical systems, positioning them not as replacements for human connection but extensions of existing support ecosystems. Models like UNICEF’s Pode Falar, which blend chatbot and offline human counselor support, illustrate how such hybrid boundary work can be deployed \cite{da2024escuta}.

\textbf{Designing for Developmental and Cultural Fit}
Adolescence is not monolithic. Our findings highlight how developmental stages shape support-seeking behavior and interaction preferences. Younger adolescents (<18) showed stronger family ties, greater vulnerability to social pressure, and higher uptake of school-based counseling. In contrast, older adolescents cited financial barriers more often, and signaled more openness towards discussions with peers. These differences suggest the need for developmentally responsive personas in chatbot design—adaptive configurations that reflect users’ age-linked needs and emotional capacities.

Culturally, many adolescents expressed frustration with the generic (often Western) orientation of MH resources. Generic advice or foreign references can erode trust. This mirrors critiques in global psychology research that stress the importance of moving beyond WEIRD (Western, Educated, Industrialized, Rich, Democratic) paradigms \cite{kozelka2021advancing, cho-etal-2023-integrative, rai2024key}. Past LMIC chatbot implementations—such as WHO’s STARS—demonstrate cultural adaptation is not a matter of translation alone \cite{who, al2024ai}. It requires rethinking tone, metaphor, structure, settings, and emotional nuance.  For instance, focus groups among Jordanian youth in the STARs project led to changing a scenario about going swimming which participants found unrelatable \cite{who}. Similarly, when the Norwegian-developed Helping Hand app was deployed among Syrian adolescents in Lebanon, many components were deemed religiously and culturally inappropriate \cite{al2021digitalized}.

\textbf{Infrastructure and the Politics of Access}
Despite high smartphone access among our sample, Indian adolescents still encounter infrastructural and social barriers that affect their engagement with MH tools. Some participants voiced discomfort using voice interfaces in shared households, as well as registration procedures that jeopardized privacy. Moreover, while our sample was primarily urban, rural adolescents may face different concerns around infrastructure (e.g., data consumption, power outages, etc.). Designing for LMIC contexts demands infrastructure-aware approaches—lightweight deployments (e.g., SMS or WhatsApp), offline fallback modes, and strong privacy features such as chat locks, local storage, or zero-data retention.

These infrastructural realities are not neutral; they reflect broader sociopolitical arrangements around digital governance, surveillance, and adolescent autonomy. A chatbot that requires a phone number for registration may unintentionally exclude those in unsafe or unsupportive homes. As such, technical decisions must be informed by ethical, political, and contextual considerations—not merely usability.

For adolescent chatbots in India to move from pilot to deployment, they must align with two national infrastructures: Tele-MANAS and the DPDP Act, 2023. Tele-MANAS (24×7 helpline, 14416) provides multilingual professional support across India; chatbots should offer clear, single-tap crisis hand-offs, preserve the user’s language choice, and supply geo-scoped referrals rather than generic links. This positions the chatbot as a bridge rather than a silo.

\textbf{Chatbots as Mediators of Care Work}
Our findings suggest chatbots are not only informational agents but also perform affective labor \cite{kerruish2021assembling}. Many participants valued when the chatbot offered validation, normalized distress, and encouraged self-reflection—roles traditionally played by trusted humans. These actions can blur the line between functional utility and emotional care.

Seen this way, chatbot design is not just a technical challenge but a form of care infrastructure design. Designers must consider how these tools coordinate care across peers, families, educators, and professionals. For example, a chatbot might scaffold a conversation between an adolescent and parent, suggest language for disclosing distress, or recommend region-specific services for in-person help.

\textbf{Toward Inclusive, Iterative, and Locally Governed Design}
To fulfill their promise, MH chatbots must be continuously co-designed with their intended users, not just during pilot phases but across their life cycle. This means building feedback loops, enabling participatory governance, and localizing both content and structure. As adolescents' needs evolve—especially across developmental transitions like school to university—chatbots must adapt alongside them.

Culturally grounded design should not imply essentializing users or assuming static identities. India’s vast heterogeneity requires plural approaches, including multilingual support, dialect-aware models, and variable chatbot personas to reflect diverse relational expectations (e.g., mentor, friend, sibling). More importantly, the chatbot must be positioned as a conversation partner, not an oracle, whose effectiveness is measured not just by retention, but by support, trust, and equity.

\textbf{Limitations and Challenges}
While our study highlights key design tensions for adolescent mental-health chatbots in India, several limitations open directions for future CHI research. First, our sample was primarily urban and female; perspectives from rural adolescents, male youth, and underprivileged groups remain underrepresented. Future work should extend this research through iterative co-design with a broader range of adolescents across India’s sociocultural and linguistic contexts.

Second, although we included adolescents under 18 in our survey, logistical and ethical constraints limited their representation in interviews. Longitudinal engagement with younger adolescents—particularly during school transitions—could deepen understanding of developmental differences in digital help-seeking.

Third, while our prototype foregrounded cultural adaptation, it lacked regional language support. Extending design to multilingual, dialect-aware chatbots will be critical for equity and uptake. Similarly, longitudinal trials are needed to assess whether adolescents sustain trust and engagement over time.

Finally, our work emphasizes chatbot design is inseparable from broader infrastructures of care. Future directions include embedding chatbots in hybrid systems that scaffold conversations with peers, parents, and counselors, while protecting adolescent privacy. Exploring participatory governance and ethical frameworks for such systems will help ensure chatbots serve not just as tools, but as inclusive infrastructures of care.

\textbf{Conclusion}
This study underscores the importance of designing digital MH tools that are  developmentally responsive and culturally grounded. By centering the voices of Indian adolescents, a historically underrepresented population in research, we reveal how stigma, infrastructural constraints, and socio-emotional needs intersect to shape their engagement with chatbots and other digital tools.

Our mixed-methods findings suggest chatbots hold significant promise as sociotechnical boundary objects—flexible, mediating artifacts adolescents interpret and use in varied ways: as confidants, advisors, sounding boards, or guides through complex care systems. Yet this potential is constrained by current tools’ lack of cultural resonance, personalization, and privacy. Adolescents expressed a need for chatbot systems that affirm emotional realities and reflect their lived environments.

More broadly, we argue culturally inclusive chatbot design requires ongoing iteration, participatory governance, and sociotechnical alignment, not simply translation or localization. Inclusive design must begin with listening—deeply, iteratively, and across boundaries—and move toward systems that are not only usable, but affirming, situated, and just. As chatbots become more embedded in adolescent lives, particularly in the Global South, their design must account for ethical, infrastructural, and contextual realities.

\bibliographystyle{ACM-Reference-Format}
\bibliography{sample-base}

@article{vogel2006measuring,
  title={Measuring the self-stigma associated with seeking psychological help.},
  author={Vogel, David L and Wade, Nathaniel G and Haake, Shawn},
  journal={Journal of counseling psychology},
  volume={53},
  number={3},
  pages={325},
  year={2006},
  publisher={American Psychological Association}
}

@inproceedings{reen2022improving,
  title={Improving mental health among working-class Indian women: insight from an interview study},
  author={Reen, Jaisheen Kour and Orji, Rita},
  booktitle={CHI Conference on Human Factors in Computing Systems Extended Abstracts},
  pages={1--6},
  year={2022}, address={New Orleans}, publisher={ACM}
}

@misc{gallup2025,
  author       = {Gallup, Inc.},
  title        = {India's Youth Dividend: High Hopes for Today and Tomorrow},
  year         = {2025},
  url          = {https://news.gallup.com/poll/509756/india-youth-dividend-high-hopes-today-tomorrow.aspx},
  note         = {Accessed: 2025-01-21}
}

@article{kozelka2021advancing,
  title={Advancing health equity in digital mental health: Lessons from medical anthropology for global mental health},
  author={Kozelka, Ellen Elizabeth and Jenkins, Janis H and Carpenter-Song, Elizabeth},
  journal={JMIR mental health},
  volume={8},
  number={8},
  pages={e28555},
  year={2021},
  publisher={JMIR Publications Toronto, Canada}
}

@inproceedings{wani2024unrest,
  title={" Unrest and trauma stays with you!": Navigating mental health and professional service-seeking in Kashmir},
  author={Wani, Asra Sakeen and Joshi, Ishika and Nahvi, Nadia Ishfaq and Singh, Pushpendra},
  booktitle={Proceedings of the 2024 CHI Conference on Human Factors in Computing Systems},
  pages={1--17},
  year={2024}, address = {Hawaii}, publisher={ACM}
}

@inproceedings{rahman2021adolescentbot,
  title={AdolescentBot: Understanding opportunities for chatbots in combating adolescent sexual and reproductive health problems in Bangladesh},
  author={Rahman, Rifat and Rahman, Md Rishadur and Tripto, Nafis Irtiza and Ali, Mohammed Eunus and Apon, Sajid Hasan and Shahriyar, Rifat},
  booktitle={Proceedings of the 2021 CHI Conference on Human Factors in Computing Systems},
  pages={1--15},
  year={2021}, address = {Remote}
}

@article{griner2006culturally,
  title={Culturally adapted mental health intervention: A meta-analytic review.},
  author={Griner, Derek and Smith, Timothy B},
  journal={Psychotherapy: Theory, research, practice, training},
  volume={43},
  number={4},
  pages={531},
  year={2006},
  publisher={Educational Publishing Foundation}
}

@article{krendl2020countries,
  title={Countries and cultural differences in the stigma of mental illness: the east--west divide},
  author={Krendl, Anne C and Pescosolido, Bernice A},
  journal={Journal of Cross-Cultural Psychology},
  volume={51},
  number={2},
  pages={149--167},
  year={2020},
  publisher={Sage Publications Sage CA: Los Angeles, CA}
}

@article{rai2024key,
  title={Key language markers of depression on social media depend on race},
  author={Rai, Sunny and Stade, Elizabeth C and Giorgi, Salvatore and Francisco, Ashley and Ungar, Lyle H and Curtis, Brenda and Guntuku, Sharath C},
  journal={Proceedings of the National Academy of Sciences},
  volume={121},
  number={14},
  pages={e2319837121},
  year={2024},
  publisher={National Acad Sciences}
}

@inproceedings{kambhamettu2024explainable,
  title={Explainable Notes: Examining How to Unlock Meaning in Medical Notes with Interactivity and Artificial Intelligence},
  author={Kambhamettu, Hita and Metaxa, Dana{\"e} and Johnson, Kevin and Head, Andrew},
  booktitle={Proceedings of the CHI Conference on Human Factors in Computing Systems},
  pages={1--19},
  year={2024},
    publisher={ACM}, address={Hawaii}
}

@article{mcdonald2019reliability,
  title={Reliability and inter-rater reliability in qualitative research: Norms and guidelines for CSCW and HCI practice},
  author={McDonald, Nora and Schoenebeck, Sarita and Forte, Andrea},
  journal={Proceedings of the ACM on human-computer interaction},
  volume={3},
  number={CSCW},
  pages={1--23},
  year={2019},
  publisher={ACM New York, NY, USA}
}

@article{murthy2017national,
  title={National mental health survey of India 2015--2016},
  author={Murthy, R Srinivasa},
  journal={Indian journal of psychiatry},
  volume={59},
  number={1},
  pages={21--26},
  year={2017},
  publisher={Medknow}
}

@misc{mental_health_survey_2022,
  title        = {Mental Health and Well-being of School Students - A Survey, 2022},
  year         = 2022,
  url          = {https://dsel.education.gov.in/node/2145},
  note         = {Accessed: 2025-01-21},
author = {Sibia, Anjum and Chakraborty, Sushmita and Shukla, Ruchi}
}

@article{lowe2005detecting,
  title={Detecting and monitoring depression with a two-item questionnaire (PHQ-2)},
  author={L{\"o}we, Bernd and Kroenke, Kurt and Gr{\"a}fe, Kerstin},
  journal={Journal of psychosomatic research},
  volume={58},
  number={2},
  pages={163--171},
  year={2005},
  publisher={Elsevier}
}

@inproceedings{cho-etal-2023-integrative,
    title = "An Integrative Survey on Mental Health Conversational Agents to Bridge Computer Science and Medical Perspectives",
    author = "Cho, Young Min  and
      Rai, Sunny  and
      Ungar, Lyle  and
      Sedoc, Jo{\~a}o  and
      Guntuku, Sharath",
    editor = "Bouamor, Houda  and
      Pino, Juan  and
      Bali, Kalika",
    booktitle = "Proceedings of the 2023 Conference on Empirical Methods in Natural Language Processing",
    month = dec,
    year = "2023",
    address = "Singapore",
    publisher = "Association for Computational Linguistics",
    url = "https://aclanthology.org/2023.emnlp-main.698/",
    doi = "10.18653/v1/2023.emnlp-main.698",
    pages = "11346--11369"
}

@article{Naveed2020PrevalenceOC,
  title={Prevalence of Common Mental Disorders in South Asia: A Systematic Review and Meta-Regression Analysis},
  author={Sadiq Naveed and Ahmed Waqas and Amna Mohyud Din Chaudhary and Sham Kumar and Noureen Abbas and Rizwan Amin and Nida Jamil and Sidra Saleem},
  journal={Frontiers in Psychiatry},
  year={2020},
  volume={11},
}

@article{patel2007mental,
  title={Mental health of young people: a global public-health challenge},
  author={Patel, Vikram and Flisher, Alan J and Hetrick, Sarah and McGorry, Patrick},
  journal={The lancet},
  volume={369},
  number={9569},
  pages={1302--1313},
  year={2007},
  publisher={Elsevier}
}

@String{Computing = "Computing" }

@String{Computer = "{IEEE} Computer" }

@article {who,
	author = {Keyan, Dharani and Hall, Jennifer and Jordan, Stewart and Watts, Sarah and Au, Teresa and Dawson, Katie S. and Sway, Rajiah Abu and Crawford, Joy and Sorsdahl, Katherine and Luitel, Nagendra P and de Graaff, Anne M. and Ghalayini, Heba and Habashneh, Rand and EL-Dardery, Hafsa and Fanatseh, Sarah and Malik, Aiysha and Servili, Chiara and Faroun, Muhannad and Abualhaija, Adnan and Aqel, Ibrahim Said and Hamdani, Syed Usman and Dardas, Latefa and Akhtar, Aemal and Bryant, Richard A. and Carswell, Kenneth},
	title = {The development of a World Health Organization transdiagnostic chatbot intervention for distressed adolescents and young adults},
	elocation-id = {2025.02.19.25322432},
	year = {2025},
	doi = {10.1101/2025.02.19.25322432},
	publisher = {Cold Spring Harbor Laboratory Press},
	URL = {https://www.medrxiv.org/content/early/2025/02/21/2025.02.19.25322432},
	eprint = {https://www.medrxiv.org/content/early/2025/02/21/2025.02.19.25322432.full.pdf},
	journal = {medRxiv}
}

@article{rupani2025like,
  title={“Like Someone Is Paying Attention to You, Listening to You, and Guiding You”: Acceptability of a Mental Health Chatbot Among Caregivers of Adolescents Living With HIV},
  author={Rupani, Neil and Vasquez, Diego H and Contreras, Carmen and Menacho, Luis and Kolevic, Lenka and Franke, Molly F and Galea, Jerome T},
  journal={Journal of the International Association of Providers of AIDS Care (JIAPAC)},
  volume={24},
  pages={23259582251327911},
  year={2025},
  publisher={SAGE Publications Sage CA: Los Angeles, CA}
}

@article{star1989institutional,
  title={Institutional ecology,translations' and boundary objects: Amateurs and professionals in Berkeley's Museum of Vertebrate Zoology, 1907-39},
  author={Star, Susan Leigh and Griesemer, James R},
  journal={Social studies of science},
  volume={19},
  number={3},
  pages={387--420},
  year={1989},
  publisher={Sage Publications London}
}

@article{national2019evidence,
  title={Evidence standards framework for digital health technologies},
  author={National Institute for Health and Care Excellence},
  journal={NICE},
  year={2019}
}

@article{world2024releases,
  title={WHO releases AI ethics and governance guidance for large multi-modal models},
  author={World Health Organization and others},
  journal={World Health Organization. Published January},
  volume={18},
  year={2024}
}

@inproceedings{daniel2022toolkit,
  title={A Toolkit for the Usability Evaluation of Digital Health Technologies},
  author={Daniel Maramba, Inocencio},
  booktitle={Proceedings of the 35th International BCS Human Computer Interaction Conference (HCI 2022)},
  year={2022}
}

@article{jha2024dpdp,
  title={THE DPDP ACT 2023: CRITICAL INSIGHTS INTO ITS LEGAL FRAMEWORK AND PRACTICAL IMPACT.},
  author={Jha, Amrita},
  journal={Bulletin of Pure \& Applied Sciences-Zoology},
  volume={43},
  year={2024}
}

@article{ahmed2022tele,
  title={Tele MANAS: India’s first 24X7 tele mental health helpline brings new hope for millions},
  author={Ahmed, Tarannum and Dumka, A and Kotwal, Atul},
  journal={Ind J Mental Health},
  volume={9},
  number={4},
  pages={403--406},
  year={2022}
}

@misc{al2024ai,
  title={AI for Health: Revolutionizing Healthcare Through Innovation},
  author={Al Mamun, Khondaker Abdullah},
  year={2024}
}

@inproceedings{lutters2002achieving,
  title={Achieving safety: A field study of boundary objects in aircraft technical support},
  author={Lutters, Wayne G and Ackerman, Mark S},
  booktitle={Proceedings of the 2002 ACM conference on Computer supported cooperative work},
  pages={266--275},
  year={2002}
}

@article{da2024escuta,
  title={A escuta sens{\'\i}vel de adolescentes atravessada por uma perspectiva docente no programa pode falar (UNICEF)},
  author={da Silva Santos, Jos{\'e} Arthur and Ferreira, Hugo Monteiro},
  journal={Debates em Educa{\c{c}}{\~a}o},
  volume={16},
  number={38},
  pages={e16574--e16574},
  year={2024}
}

@inproceedings{zhou2011cpoe,
  title={CPOE workarounds, boundary objects, and assemblages},
  author={Zhou, Xiaomu and Ackerman, Mark and Zheng, Kai},
  booktitle={Proceedings of the SIGCHI Conference on Human Factors in Computing Systems},
  pages={3353--3362},
  year={2011}
}

@mastersthesis{al2021digitalized,
  title={Digitalized Psychosocial Support in Education: exploring the impact of the happy helping hand app for displaced syrian adolescents in Lebanon},
  author={Al-Khayat, Alaa Munir},
  year={2021},
  school={OsloMet-Storbyuniversitetet}
}

@article{yoo2024missed,
  title={Missed opportunities for human-centered AI research: Understanding stakeholder collaboration in mental health AI research},
  author={Yoo, Dong Whi and Woo, Hayoung and Pendse, Sachin R and Lu, Nathaniel Young and Birnbaum, Michael L and Abowd, Gregory D and De Choudhury, Munmun},
  journal={Proceedings of the ACM on Human-Computer Interaction},
  volume={8},
  number={CSCW1},
  pages={1--24},
  year={2024},
  publisher={ACM New York, NY, USA}
}

@article{kerruish2021assembling,
  title={Assembling human empathy towards care robots: the human labor of robot sociality},
  author={Kerruish, Erika},
  journal={Emotion, Space and Society},
  volume={41},
  pages={100840},
  year={2021},
  publisher={Elsevier}
}

@misc{sehgal2025paldesigningconversationalagents,
      title={PAL: Designing Conversational Agents as Scalable, Cooperative Patient Simulators for Palliative-Care Training}, 
      author={Neil K. R. Sehgal and Hita Kambhamettu and Allen Chang and Andrew Zhu and Lyle Ungar and Sharath Chandra Guntuku},
      year={2025},
      eprint={2507.02122},
      archivePrefix={arXiv},
      primaryClass={cs.HC},
      url={https://arxiv.org/abs/2507.02122}, 
}

\clearpage
\onecolumn
\appendix

\section{Extended Methodology}

\textbf{Study Design}
This study employed a mixed-methods approach to explore the mental health needs, barriers, and preferences of adolescents in India. The research was conducted in two phases: (1) a quantitative survey to capture broad trends and (2) in-depth qualitative interviews to gather rich, contextual insights. The survey provided a high-level understanding of participants’ experiences and attitudes, while the interviews allowed us to dive deeper into the nuances behind the quantitative findings. This approach allowed us to triangulate findings and develop comprehensive recommendations for the design of culturally inclusive digital mental health tools.

\textbf{Participant Recruitment}
Participants were recruited through a combination of online platforms, including social media and outreach to school administrators. We aimed to target participants from a range of geographic regions and socioeconomic backgrounds. Upon completion of the anonymous surveys, participants were redirected to a second survey, asking if they were interested in participating in an in-depth interview. All participants completing the survey were entered into a raffle for one of ten 2000 INR gift cards, and all interview participants received a 500 INR gift card.

\textbf{Phase 1: Survey}
The survey was designed to collect quantitative data on participants’ mental health experiences, perceived barriers to seeking support, and preferences for digital mental health tools. The survey consisted of both closed-ended questions (e.g., Likert scales and multiple-choice) and open-text fields for additional comments. Questions were developed with the assistance of mental health counselors in India and were also drawn from previously validated surveys and scales \cite{vogel2006measuring, murthy2017national, mental_health_survey_2022, lowe2005detecting}. Responses were collected via Qualtrics and all questions were in English.

\textbf{Phase 2: Qualitative Interviews}
To gain deeper insights into the themes identified in the survey, we conducted semi-structured interviews with 12 participants. The interviews primarily consisted of a role-play activity, along with in-depth questions about participants’ mental health experiences and preferences. 

The role-play activity was designed to simulate real-life challenges commonly faced by adolescents in India, based on discussions with mental health counselors. Participants were asked to choose one of six scenarios (see in Appendix section C) reflecting these challenges and were asked to engage in two tasks:

\begin{itemize}
    \item \textbf{Searching for Information Online}: Participants were asked to search for information or solutions to the scenario using any online platform they felt comfortable with (e.g., Google, YouTube Shorts, Instagram, Reddit). They were instructed not to use chatbots for this task. Participants shared their screens during this activity, allowing us to observe their search behavior and the resources they consulted.

    \item \textbf{Interacting with a Custom Chatbot}: Participants then engaged with a basic chatbot prototype we developed for the study (Figure 1). The chatbot was built on top of GPT-4o-mini and designed to be a supportive and empathetic conversational agent aimed at helping Indian students facing challenges. It was not intended to provide therapy or replace professional counselors but to offer understanding, encouragement, and practical suggestions. The chatbot’s responses were structured around a predefined prompt developed in collaboration with mental health counselors in India. The full prompt is provided in the Appendix section D.

\end{itemize}

Interviews were conducted via Zoom in English, with each session lasting approximately 45-60 minutes. This combination of role-play and qualitative inquiry provided rich, contextual insights into participants’ mental health needs, barriers, and preferences for digital mental health tools.

\textbf{Data Analysis}
Survey data were analyzed using descriptive statistics. After computing descriptive statistics, we dichotomised respondents by chronological age (<18 years vs. $\geq$ 18 years). Single‑choice survey items were tested for age differences with 2xK Pearson ${\chi}^2$ tables and reported with Cramer’s V. For each “check‑all‑that‑apply” question we split selections into binary dummies and ran 2x2 ${\chi}^2$ tests, reporting $\phi$. A threshold of p<.01 was adopted to guard against family‑wise error across all tests.

Thematic analysis was employed to examine the interview transcripts. One author initiated the process by performing an open-coding pass to generate an initial set of codes. These codes were subsequently reviewed by a second author, who collaborated with the first author to refine them. This refinement involved discarding codes that were not instrumental in capturing participants' needs or proposing enhancements to digital mental health tools. To ensure validity, detailed discussions were conducted throughout each phase of the analysis, emphasizing a collaborative review to maintain accuracy and consistency in coding. This method was preferred over calculating inter-rater reliability (IRR), as the intricate nature of the codes and the context-sensitive discussions provided deeper insights \cite{mcdonald2019reliability, kambhamettu2024explainable, sehgal2025paldesigningconversationalagents}.

\textbf{Ethics}
The study was approved by our institutional review board, and all participants provided informed consent prior to participation. Survey responses were anonymous, and interviews were recorded with explicit consent.


\section{Participant Demographics}

\begin{table}[H]
\centering
\caption{Self Reported Demographics}
\label{tab:participant-info}
\begin{tabular}{@{}llc@{}}

\textbf{Participant ID} & \textbf{Age} & \textbf{Gender} \\

P1 & 20 & Female \\
P2 & 21 & Female \\
P3 & 19 & Female \\
P4 & 18 & Female \\
P5 & 18 & Female \\
P6 & 20 & Female \\
P7 & 20 & Female \\
P8 & 18 & Male \\
P9 & 19 & Female \\
P10 & 19 & Female \\
P11 & 14 & Male \\
P12 & 18 & Female \\

 P13& 18&Female\\
 P14& 20&Male\\
\end{tabular}
\end{table}

\lstset{
    basicstyle=\ttfamily\small,
    frame=single,
    breaklines=true,
    columns=fullflexible,
    numbers=left,
    numberstyle=\tiny,
    stepnumber=1,
    showspaces=false,
    showstringspaces=false,
    breakatwhitespace=true,
}

\section{Roleplay Scenarios}

\begin{lstlisting}
Scenario A: Struggling to Get an Entrance 
You've been preparing for university entrance exams for years, but despite your efforts, you didn't achieve the score you needed to secure admission to your dream college. You feel like you've let your family down and are losing confidence in yourself. Your parents have invested so much in your education, and you're worried they'll see you as a failure. How can you cope with these feelings and move forward?

Scenario B: Financial Issues and Guilt for Parents' Support
Your family is facing significant financial challenges, and you're acutely aware of the sacrifices your parents have made to support your education and dreams. You feel a heavy sense of guilt every time you ask for something or even when you spend money on basic needs. This guilt is affecting your motivation and your relationship with your family. How can you deal with these emotions and focus on your goals?

Scenario C: Problems in Love
You're in a romantic relationship, but it has become increasingly complicated. You often feel unsupported or misunderstood, and arguments have become frequent. You're torn between trying to fix things and ending the relationship, but you're afraid of the judgment and stigma that might come with discussing your relationship struggles openly. What should you do to address these feelings?

Scenario D: Abuse (Physical/Emotional/Verbal)
Someone close to you-whether at home, school, or in a relationship-has been treating you in a way that feels abusive or harmful. This might include harsh words, controlling behavior, or even physical harm. You're scared to speak up because you're not sure if anyone will believe you or how they might react. How can you find a safe way to talk about this and get the help you need?

Scenario E: Fitting In Socially:
You recently moved to a new school and are struggling to make friends. You feel lonely and don't know how to start a conversation with your classmates. What are some ways to build friendships and feel more connected?

Scenario F: Loneliness and Isolation
You feel like no one understands you or cares about how you feel. Even when you're around others, you feel lonely, and you've started avoiding social situations. What can you do to feel less isolated?

\end{lstlisting}

\section{Chatbot System Prompt}

\begin{lstlisting}
You are a supportive and empathetic chatbot designed to help Indian high school students who are facing challenges. Your role is not to provide therapy or replace professional counselors, but to:
- Offer understanding and encouragement.
- Ask questions to better understand the user's situation before offering suggestions.
- Provide practical suggestions or coping strategies only after understanding their problem.
- Encourage healthy behaviors like talking to a trusted adult, seeking professional help, or engaging in self-care.
- Avoid making diagnoses or offering medical advice.
- Acknowledge cultural nuances and the pressures that Indian high schoolers face, such as academic expectations, social dynamics, and family relationships.
- Always prioritize the safety and well-being of the user.

When responding:
1. Be warm, understanding, and non-judgmental.
2. Start by acknowledging the user's feelings and asking them to share more about their situation.
3. Ask for background details on the user's situation.
4. Only offer suggestions or next steps after understanding their concerns.
5. Focus on empowering the user by normalizing their feelings and validating their concerns.
6. Suggest accessible resources or actionable next steps that could realistically help them.
7. If the problem described suggests a need for professional mental health care, gently encourage the user to consider reaching out to a counselor, psychologist, or school support staff.

Examples of phrases you can use:
- "That sounds really tough; would you like to share more about what's been going on?"
- "I'm here to listen; can you tell me a little more about how you're feeling?"
- "It's completely normal to feel this way. Could you share what's been weighing on you the most?"
- "You're not alone in feeling this way. Could you tell me more about what's happening?"
- "Once I understand a bit more, I'd be happy to suggest something that might help."

Always keep your tone supportive and avoid overwhelming the user with too much information at once.
- Always give short responses: 2-3 sentences, less than 50 words.
- Ask one question per turn to keep the conversation focused.
- Use separate lines for responses and questions.
- Don't encourage the user to speak to someone in every turn, but do suggest it when appropriate.
- Prioritize asking for details to better understand the user's problem.
\end{lstlisting}


\end{document}